\documentclass[]{aa}  

\usepackage{graphicx}
\usepackage{txfonts}
\usepackage{xcolor}
\usepackage{hyperref}
\newcommand{\teff}{T$_{\rm eff}$ }
\newcommand{\logg}{$\log(g)$ }

\newcommand{\vmic}{$\xi_{\rm mic}$ }

\begin{document} 

   \title{Abundances of disk and bulge giants from \\
high-resolution optical spectra \thanks{Based on observations made with the Nordic Optical Telescope (programs 51-018 and 53-002) operated by the Nordic Optical Telescope Scientific Association at the Observatorio del Roque de los Muchachos, La Palma, Spain, of the Instituto de Astrofisica de Canarias, spectral data retrieved from PolarBase at Observatoire Midi Pyrénées, and observations collected at the European Southern Observatory, Chile (ESO programs 71.B-0617(A), 073.B-0074(A), and 085.B-0552(A)).}}

  \subtitle{V. Molybdenum – the p-process element}

   \author{R. Forsberg
          \inst{1}
          \and
          N. Ryde\inst{1} 
          \and 
          H. Jönsson\inst{2}
          \and
          R. M. Rich\inst{3}
          \and
          A. Johansen \inst{1,4}
          }

   \institute{Lund Observatory, Department of Astronomy and Theoretical Physics, Lund University, Box 43, SE-22100 Lund, Sweden 
   \\ 
              \email{rebecca@astro.lu.se}
    \and 
    Materials Science and Applied Mathematics, Malm\"o University, SE-205 06 Malm\"o, Sweden 
    \and         
    Department of Physics and Astronomy, ICLA, 430 Portola Plaza, Box 951547, Los Angeles, CA 90095-1547 
    \and 
    Center for Star and Planet Formation, GLOBE Institute University of Copenhagen, Øster Voldgade 5-7, 1350 Copenhagen, \newline Denmark 
    }

   \date{Received Month xx, xxxx; accepted Month XX, XXXX}

 
  \abstract
{}
{In this work, we aim to make a differential comparison of the neutron-capture and p-process element molybdenum (Mo) in the stellar populations in the local disk(s) and the bulge, focusing on minimising possible systematic effects in the analysis.}
{The stellar sample consists of 45 bulge and 291 local disk K-giants, observed with high-resolution optical spectra. The abundances are determined by fitting synthetic spectra using the SME-code. The disk sample is separated into thin- and thick-disk components using a combination of abundances and kinematics. The cosmic origin of Mo is investigated and discussed by comparing with previous published abundances of Mo and the neutron-capture elements cerium (Ce) and europium (Eu).}
{We determine reliable Mo abundances for 35 bulge and 282 disk giants with a typical uncertainty of [Mo/Fe] $\sim 0.2$ and $\sim 0.1$ dex for the bulge and disk, respectively.}
{We find that the bulge possibly is enhanced in [Mo/Fe] compared to the thick disk, which we do not observe in either [Ce/Fe] nor [Eu/Fe]. This might suggest a higher past star-formation rate in the bulge, however, since we do not observe the bulge to be enhanced in [Eu/Fe], the origin of the molybdenum enhancement is yet to be constrained. Although, the scatter is large, we may be observing evidence of the p-process contributing to the heavy element production in the chemical evolution of the bulge.}


   \maketitle
%

\defcitealias{jonssona2017A&A...598A.100J}{Paper I}
\defcitealias{jonsson2017b}{Paper II}
\defcitealias{lomaeva2019A&A...625A.141L}{Paper III}
\defcitealias{Forsberg2019A&A...631A.113F}{Paper IV}

\section{Introduction}
Elemental abundances of stars has proven key both in tracing the chemical evolution of the Milky Way and to understand the origin of the elements themselves. Stars carry a chemical fingerprint from the molecular cloud from which they formed, which can be measured in their photospheres. The Galaxy is enriched with elements over time, where they are formed in various processes, either internally in stars, or in more explosive environment such as supernovae (SNe) type Ia and II and neutron star mergers (NSM).

By measuring the chemical abundance in stars of a range of metallicities, [Fe/H], we can trace the evolution of elements and in turn, the stellar populations that make up the Milky Way. In this work we focus on the disk components (thin and thick) and the bulge. The origin and evolution of the bulge have attracted much attention, and the bulge has been re-defined from a classical, spherical bulge to now being primarily classified as a psuedo-bulge with a box/peanut bar \citep[e.g.][]{ness2012ApJ...756...22N,dimatteo2014A&A...567A.122D,shen2020RAA....20..159S}. The connection of the bulge/bar to the disk structure remains central, and careful, detailed abundance studies show surprisingly small differences between the composition of the bulge and the thick disk \citep{jonsson2017b,lomaeva2019A&A...625A.141L,Forsberg2019A&A...631A.113F}. While it is agreed that the bulge metallicities reaches values higher than the thick disk \citep{matteucci1990ApJ...365..539M,mcwilliam2016PASA...33...40M}, there remains no convincing difference in abundance trends \citep[see e.g.][and references therein]{barbuy2018ARA&A..56..223B}. A careful differential comparison between the bulge and the disk is needed, and while many such studies are published, ours employs newly reduced and analyzed high quality datasets.  

In this article series, \citep[][hereafter refereed to as \citetalias{jonssona2017A&A...598A.100J},\citetalias{jonsson2017b},\citetalias{lomaeva2019A&A...625A.141L}, \citetalias{Forsberg2019A&A...631A.113F}]{jonssona2017A&A...598A.100J,jonsson2017b,lomaeva2019A&A...625A.141L,Forsberg2019A&A...631A.113F} the bulge chemistry is investigated by a differential comparison to a disk sample. These have been analyzed with the same method, atomic data and set of spectral lines, from high-resolution spectra of 291 disk giants and 45 bulge giants. This has been done in order to minimize the systematic uncertainties in the analysis, especially by using the same type of stars which removes the possible systematic difference in abundance between dwarf and giant stars \citep[see e.g.][]{melendez2008A&A...484L..21M,gonzalez2016ASSL..418..199G}. In the previous papers we find that the bulge has a similar evolutionary history as the thick disk, although we can not exclude possible relative enrichment in some elements, like vanadium (V), cobalt (Co) \citepalias{lomaeva2019A&A...625A.141L}, and lanthanum (La) \citepalias{Forsberg2019A&A...631A.113F} in the bulge.

In \citetalias{Forsberg2019A&A...631A.113F} we investigated the neutron-capture elements Zr, La, Ce and Eu. These are produced through neutron-capture processes \citep{cameron1957PASP...69..201C,burbidge1957}, which is the process responsible for creating more than two-thirds of the periodic table of elements. These can add an additional piece to the puzzle in Galactic archaeology, different from both the $\alpha$- and iron-peak elements which have primarily SNe type II and SNe type Ia origin.

The neutron-capture process involves two physical processes. The capture of a neutron onto a seed atom, creating a heavier isotope, and the possible subsequent $\beta^-\mathrm{-decay}$, $\mathrm{n} \to \mathrm{p} + \mathrm{e}^- + \nu_\mathrm{e}$, creating a heavier element. As a consequence, the neutron-capture and $\beta^-$-decay outline two sub-processes, the \textit{slow} s-process, and the \textit{rapid} r-process.

While it is usual to refer to elements as either primarily being either s- or the r-process, in virtually all cases, both r- and s-process contribute. An r-process element is an element with a \textit{dominant} origin from the r-process in the Sun, and vice versa for s-process elements. It can be more informative to examine the (stable) isotopes that make up the element, since it all comes down to how those elements are formed. For instance, europium (Eu, Z = 63) has two stable isotopes that contribute roughly 50/50 to the solar Eu, ${}^{152}\mathrm{Eu}$ and ${}^{153}\mathrm{Eu}$ \citep{bisterzo2014ApJ...787...10B,prantzos020MNRAS.491.1832P}, where both of them have a dominating origin from the r-process, making Eu an r-process element.

35 stable isotopes cannot be reached through either the s- or r-process and are instead formed in the so-called p-process \citep{cameron1957PASP...69..201C,burbidge1957}, which will be discussed more in Section\,\ref{sec: discussion}. 

The element molybdenum\footnote{Molybdenum is an element both discovered and first time isolated by the Swedish chemists Carl Wilhelm Scheele and Peter Jacob Hjelm in the 18th century.} (Mo, Z = 42) has seven stable isotopes: ${}^{92,94,95,96,97,98,100}\mathrm{Mo}$. Mo is an intriguing element since it has pure isotopes from both the s-, r- and the p-processes. Both of the lightest ones, ${}^{92}\mathrm{Mo}$ and ${}^{94}\mathrm{Mo}$, are p-isotopes, whereas ${}^{96}\mathrm{Mo}$ is a pure s-isotope and ${}^{100}\mathrm{Mo}$ is a pure r-isotope. Combined, the p-isotopes contribute to roughly $20-25\ \%$ of the solar Mo abundance, which is the highest contribution from the p-process seen in any element \citep[as measured in the Sun,][]{prantzos020MNRAS.491.1832P}. The second highest p-element is ruthenium (Ru, Z = 44) with a roughly $7\ \%$ p-process contribution in the Sun \citep{prantzos020MNRAS.491.1832P}. As such, the Mo and its isotopes provide excellent benchmark examples for nuclear- and astrophysics. Studying Mo from a Galactic chemical evolution (GCE) perspective can help to put constraints on the origin of this element, the origin of the p-process, and to constrain the evolution of the Galaxy itself.  

The remaining contributions from the s- and r-process to Mo have varying values, where \citet{bisterzo2014ApJ...787...10B} report Mo to have a $39\%$ production from the s-process, whereas \citet{prantzos020MNRAS.491.1832P} instead give a $50\%$ s-process and $27\ \%$ r-process origin to Mo, at solar metallicities. 

However, it should be noted that the solar composition of molybdenum might not be representative of the Galactic composition often presented in Galactochemical evolution models. Measurements of meteorites show that the \textit{inner} Solar System, especially the Earth, is relatively enriched in the s-process by high ${}^{96}\mathrm{Mo}$ values, compared to ${}^{92,94}\mathrm{Mo}$ (p-process) and ${}^{95}\mathrm{Mo}$ (r-process dominated) \citep{burkhardt2011E&PSL.312..390B,budde2016E&PSL.454..293B,budde2019NatAs...3..736B}. This way the Mo isotopic composition of Earth can be used to make constraints on the types of meteoric material that contributed to the formation of our planet. The meteoric isotopic composition has been suggested to be affected by where the dust, which make up the meteorites, originated from \citep{lugaro2016PNAS..113..907L,ek2020NatAs...4..273E}. This makes Mo particularly interesting to study and to put further constraints on its cosmic origin. 

Larger published samples of Mo abundances for stars in the disk and bulge are sparse. In the work of \citet{mishenina2019MNRAS.489.1697M} they determine Mo-abundances for roughly 200 dwarf disk stars, which complements previous abundances at metallicities of [Fe/H] $< - 1.2 \ \mathrm{dex}$ \citep{peterson2013ApJ...768L..13P,roederer2014...313...AJ....147..136R,hansen2014A&A...568A..47H,spite2018A&A...611A..30S}. However, in Galactic chemical evolution models, Mo has been consequently underestimated compared to observations \citep{mishenina2019MNRAS.489.1697M}. On the other hand, in the models in \citet{kobayashi2020ApJ...900..179K} where they do include $\nu$-winds (which is a suggested production channel for p-isotopes), they overproduce Mo, indicating that it is a problematic element to constrain without proper knowledge and modeling of the p-process. By presenting abundances for Mo in both giants in the local disk and in the bulge, we aim to put further constraints on both the Galactochemical evolution of the bulge and to the origin of Mo.

This paper is structured as follows: in Section\,\ref{sec: Data} we present the spectroscopic data used. In Section\,\ref{sec: methodology} we present the methodology for the analysis of the data, which follows closely that of previous papers in this series. In Section\,\ref{sec: results} we present the abundances, the estimated uncertainties and comparison with previous studies. Finally, in Section \,\ref{sec: discussion} and \ref{sec: conclusions} we discuss the results and conclude our findings. 

\section{Data}
\label{sec: Data}
In this section we introduce the data used in this work, where we aim to have high resolution and high signal-to-noise (S/N) spectra for our bulge- and disk giants. 

\subsection{Bulge}
\begin{figure*}
   \centering
\includegraphics[width=\hsize]{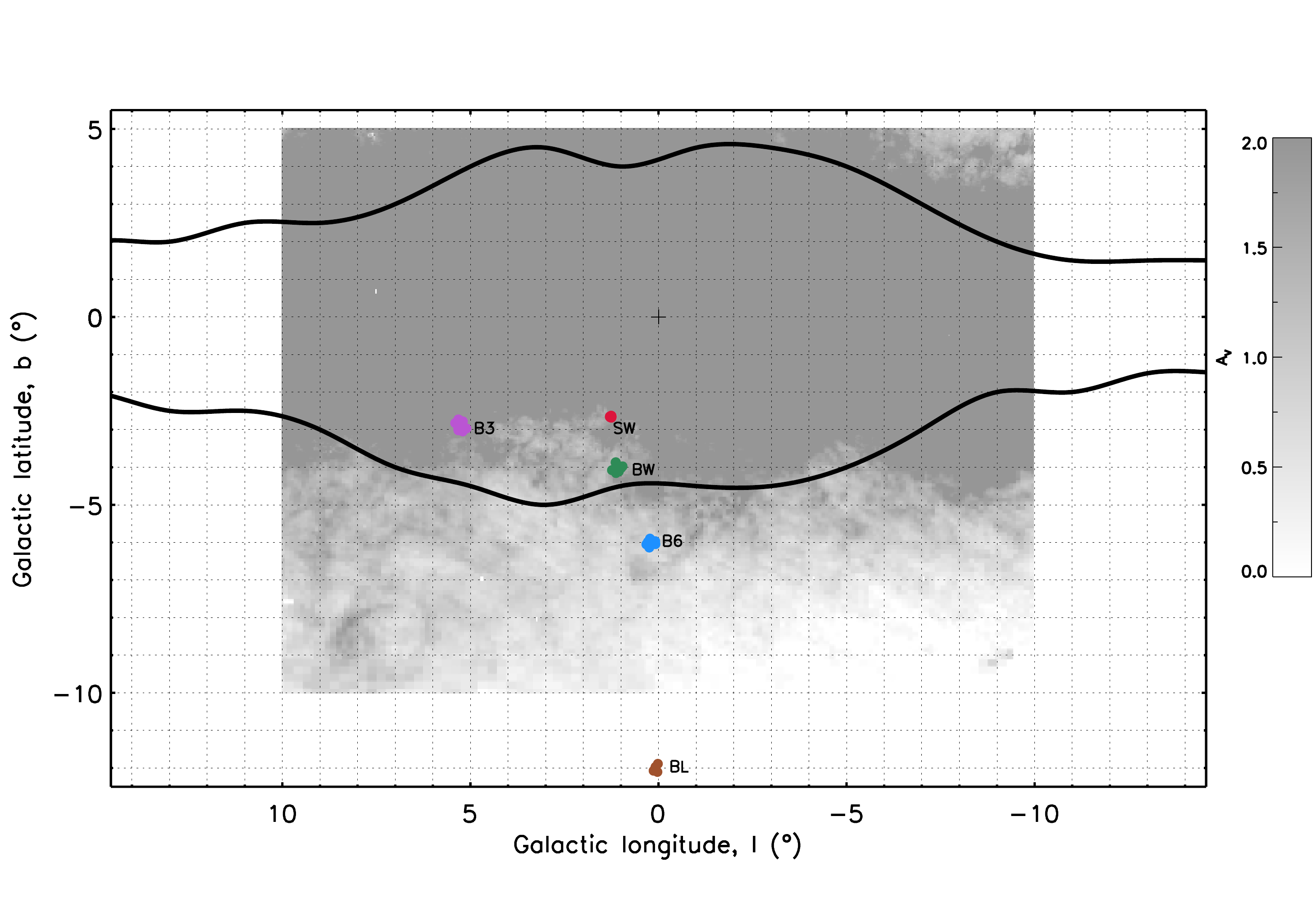}
 \caption{Map of the Galactic bulge showing the five analysed fields, B3, BW, B6, BL \citep[using the naming convention in ][]{lecureur2007} and SW. The dust extinction towards the bulge is taken from \citet{gonzalez:11,gonzalez2012} and scaled to optical extinction \citep{cardelli}. The scale saturates at A$_{\text{V}} = 2$, which is the upper limit in the figure. The COBE/DIRBE contours of the Galactic bulge, in black, are from \citet{weiland:94}.}
\label{fig: bulge fields}
    \end{figure*}

Optical high-resolution spectra of bulge giants are fairly rare, given by the long integration times needed for observing these stars. Additionally, to observe bulge giants in the optical wavelength regime is a challenge in itself, given the high amount of dust causing extinction. The spectra used here have thus been collected from low extinction regions in the bulge, or its vicinity, which can be seen in Figure\,\ref{fig: bulge fields}

The spectra for stars in the B3, BW, B6, BL fields \citep[using the naming convention in][]{lecureur2007} are obtained in 2003-2004 whilst the SW field was obtained in 2011 (ESO program 085.B-0552(A)). They are all obtained with the UVES/FLAMES spectrograph ($R \sim 47 000$) mounted on the VLT and are limited to the wavelength regime of 5800-6800 Å. The signal-to-noise ratio \citepalias[see][for details of the S/N estimation]{jonssona2017A&A...598A.100J} are generally around 50, see Table\,\ref{tab:basicdata_bulge} for details of all the bulge giants and their spectra.

The spectra from the B3, BW, B6, BL fields were first used in \citet{zoccali2006A&A...457L...1Z} and were reanalyzed in several subsequent articles, such as \citet{lecureur2007,vanderswaelmen}. In \citetalias{jonsson2017b, lomaeva2019A&A...625A.141L,Forsberg2019A&A...631A.113F} we re-analyze 27 of these bulge stars, plus the additional 18 stars in the low-extinction SW field, which adds up to the 45 stars we have analysed in this paper. The reader is refereed to \citetalias{jonsson2017b} for further details of the bulge sample.

\subsection{Disk}
The disk sample consists of 291 giant local disk stars. 272 of these are observed by us using the FIbre-fed Echelle Spectrograph \citep[FIES][]{telting:2014}, mounted on the Nordic Optical Telescope, Roque de Los Muchachos, La Palma and 19 spectra are downloaded from the PolarBase data base \citep{petit:2014}, in turn coming from the ESPaDOnS and NARVAL spectrographs (mounted on Canada-France-Hawaii Telescope and Telescope Bernard Lyot, respectively). The FIES and PolarBase spectrographs have similar resolutions of $R \sim 67 000$ and $R \sim 65 000$, respectively, and wavelength coverage of 3700-8300 Å and 3700-10500 Å, respectively. However, note that we only use the 5800-6800 Å wavelength regime, to match that of the bulge spectra and to only use the same spectral lines in the analysis. The FIES spectra have a S/N of around 80-120, whereas the PolarBase are lower, around 30-50. All spectra are reduced using the standard automatic pipelines. See Table\,\ref{tab:basicdata_sn} for details of all the disk giants and their spectra. 

A telluric spectrum, namely the one in the Arcturus atlas \citep{arcturusatlas}, has been plotted over the observed stellar spectra such that regions affected by telluric lines could be avoided on a star-by-star basis. Further details of the FIES observational programs and the disk spectra are found in \citetalias{jonssona2017A&A...598A.100J}. 

\section{Methodology}
\label{sec: methodology}
The methodology of the analysis in this work strongly follows the methodology as set out in the previous papers in this series; \citetalias{jonssona2017A&A...598A.100J,jonsson2017b,lomaeva2019A&A...625A.141L,Forsberg2019A&A...631A.113F}. In those previous papers, we obtain very tight abundance trends with [Fe/H], and we can see that a carefully chosen set of spectral lines are key to achieving these high quality abundances. In this section, we will go trough the basics of the analysis of the giant stars, especially focusing on the 6030 Å Mo I line used in this analysis. 

\subsection{Spectral analysis}
The spectral analysis to obtain the stellar parameters and elemental abundances is done using the tool Spectroscopy Made Easy \citep[SME, version 554]{valentisme1996,piskunovSME2017}. SME produces a synthetic spectrum using a $\chi^2$-minimization to fit towards the observed spectrum. 
To produce a synthetic spectrum, SME requires:
\begin{itemize}
    \item[–] A line list containing atomic- and/or molecular data. We use the Gaia-ESO line list version 6 as published in \citet{heiter2021A&A...645A.106H}.
    
    \item[–] Model atmospheres; in this work use the grid of MARCS models\footnote{Available at \url{(marcs.astro.uu.se)}} \citet{gustafsson2008}. Since our stellar sample consists of giant stars, we use the MARCS models with spherical symmetry for \logg $< 3.5$. 
    
    \item[–] Stellar parameters; the ones used in this work are derived in \citetalias{jonssona2017A&A...598A.100J,jonsson2017b}, where more details can be found. In short, we use a combination of Fe I and Fe II lines, Ca I lines, and $\log g$ sensitive Ca I line wings. The Fe I lines have NLTE corrections adopted from \citet{lind2012:nlte}. In \citetalias{jonssona2017A&A...598A.100J,jonsson2017b} we also estimate typical uncertainties and compare with Gaia benchmark stars \citep{heiter2015A&A...582A..49H,jofre2014A&A...564A.133J,jofre:2015}. In general, the stellar parameters compare well with the benchmark parameters. The surface gravitites are however likely systematically high from the benchmark comparison with +0.10 dex. Furthermore, comparing with StarHorse \citep{queiroz2018} surface gravities derived from Gaia EDR3 \citep{gaiacollaboration2016A&A...595A...1G,gaiacollaboration2021A&A...649A...1G,anders2022}, the \citetalias{jonssona2017A&A...598A.100J} surface gravities are as well +0.10 dex too high, systematically. This could cause some overestimation of the abundances, however the main scope of this work is to make a differential comparison between the disk and bulge sample, that will be equally systematically affected.
    
    \item[–] A defined spectral segment, within which the line of interest and local continuum is marked with a line mask, or continuum masks, respectively. By the manual placement of local continuum masks, the continuum is renormalised\footnote{A crude normalisation has already been done using the IRAF \texttt{continuum} tool.} more carefully to the local segment around the spectral line of interest. SME uses the continuum masks to fit a straight line inbetween, creating the local continua.
    
    The line mask around the line of interest, the Mo I 6030 Å line in this case, is also defined manually. This manual placement of both line- and continuum masks has been shown crucial in order to get high-precision abundances \citepalias{jonssona2017A&A...598A.100J,jonsson2017b,lomaeva2019A&A...625A.141L,Forsberg2019A&A...631A.113F}. The line- and continuum masks can be seen in Figure\,\ref{fig: continuum and line segments of Mo,both}. We go into more details of the abundance determination below. 
\end{itemize}

\subsection{Abundance determination of Mo}
\label{sec: abu det Mo}
In the abundance determination we use the Mo I spectral line located at 6030 Å. The atomic data we use for this line comes from the Gaia-ESO line list version 6 \citep{heiter2021A&A...645A.106H}, see Table\,\ref{tab: atomic data}. The spectral line is classified as a \textit{Yes}/\textit{Yes} line in the Gaia-ESO list, meaning that it both has a high quality $\log(gf)$-value and is unblended. Furthermore, in \citet{heiter2021A&A...645A.106H} they report that the line should be avoided in abundance analysis of dwarf stars. As such, this line is a great example of a line that can only be reached in giant stars, where the lower surface gravities increase the line strength sufficiently enough for reliable abundances to be estimated. 

Mo does not have any hyperfine splitting (HFS) but has, as before mentioned, seven stable isotopes in the Sun. However, these are not included in the line list since the isotopic shift (IS) can not be resolved, partly due to the Mo I lines being very weak. This means that we can not measure the individual isotopic abundances, but rather the molybdenum abundance as a whole.

    \begin{table}
    \caption{Atomic data for the Mo I atomic line. The wavelength is the wavelength in air. Data from \citet{WBb1988PhyS...38..707W}, as compiled in the Gaia-ESO. v.6 line list \citet{heiter2021A&A...645A.106H}.}  
    \label{tab: atomic data}  
    \centering     
    \begin{tabular}{c | c | c | c }  
    \hline\hline     
    Element & Air wavelength [Å] & $\log(gf)$ & $\chi_{\text{ exc}}^{\text{ low}}$ [eV] \\
    \hline \hline 
    Mo I & 6030.644 & -0.523 & 1.531 \\
    \hline   
    \end{tabular}
    \end{table}

The observed and synthetic spectra close to the 6030 Å line can be seen in Figure\,\ref{fig: continuum and line segments of Mo,both} for both a typical, bright red giant disk star (Arcturus/$\alpha$-Boo/HIP69673, in the top row) and one of the bulge stars (B3-B8, in the bottom row). Here we also plot the line- and continuum segments used to produce the synthetic spectra. The same masks are used for all stars in the analysis, to be as coherent as possible. 

All synthetic spectra are examined by eye and and the masks edited to return the best possible fit of the synthetic spectra over the observed spectrum. For the stars where a line is detectable and above the noise, reliable synthetic spectra, and in turn, abundances, could be determined. In instances where the spectral line is weaker than the noise, the line can not be used to determine a reliable abundance. Nonetheless, using giant stars we are able to determine the abundance of low molybdenum abundance stars, since the line strengths typically increases with decreasing surface gravity.
It should be noted that we have determined all Mo abundances under the assumption of LTE. The 6030 Å line is a rather weak line, and forms in the deeper parts of the stellar atmosphere where collisions dominate, establishing LTE. As such, NLTE corrections for Mo should be small or negligible, which also has been noted in previous studies with smaller sample of stars \citep[see e.g. ][]{peterson2011ApJ...742...21P,roederer2014ApJ...791...32R,roederer2022arXiv220503426R}. 

  \begin{figure*}[h!]
   \centering
    \includegraphics[width=\hsize]{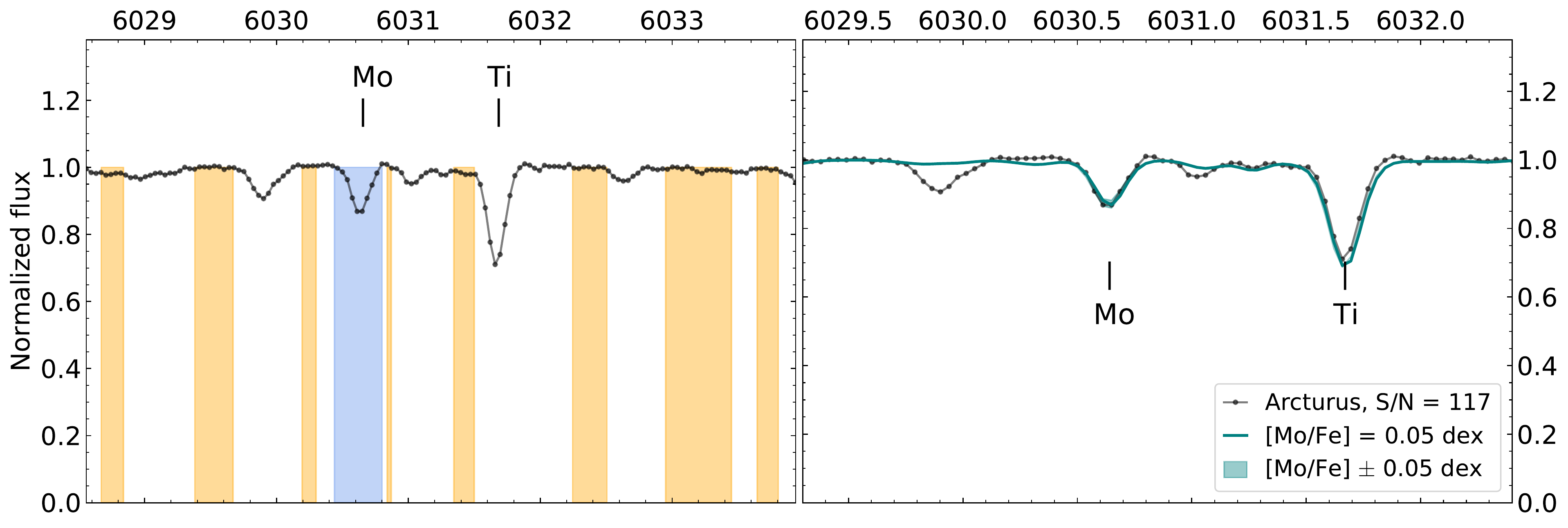}
    \includegraphics[width=\hsize]{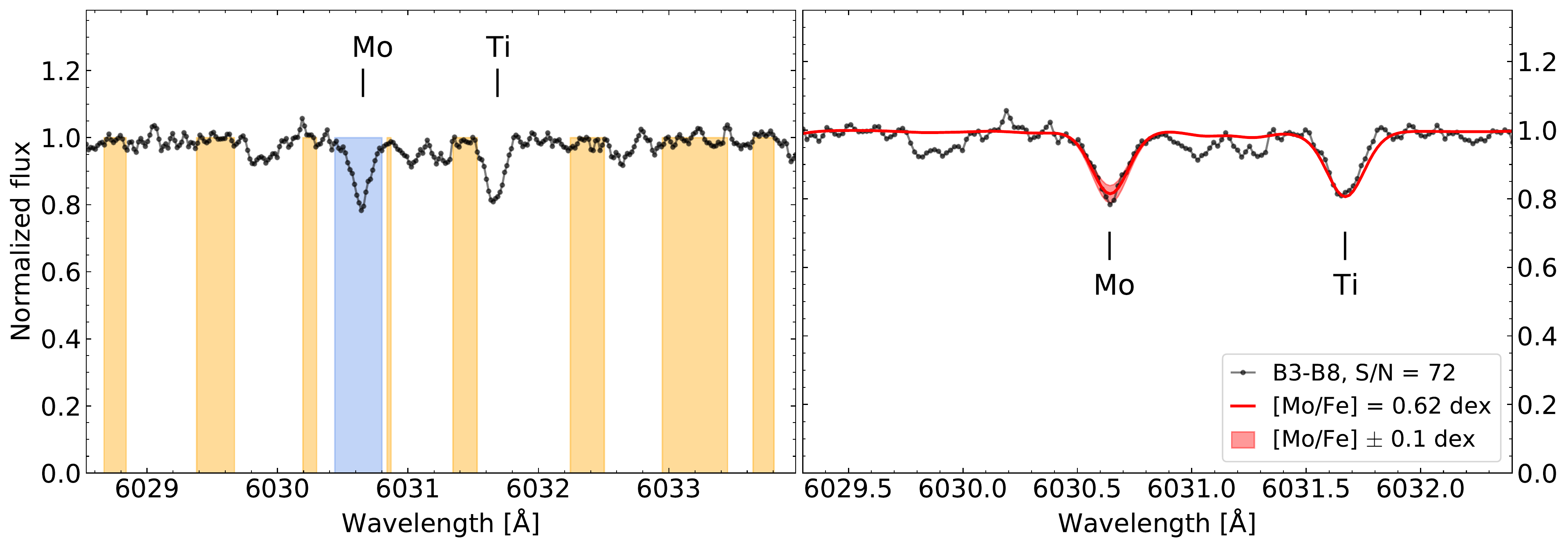}
      \caption{The observed spectrum (black) for a disk star (top, Arcturus) and bulge star (bottom, B3-B8). Left: The line mask placements for the Mo line can be seen in blue and the continuum placements in yellow. Right: The synthetic spectra, either in teal (top, disk star Arcturus) or red (bottom, bulge star B3-B8). The estimated S/N of both spectra, 117 and 72 respectively, is indicated in the legend. Note that the wavelength region of the rightmost figures are zoomed in with respect to the leftmost figures.}
      \label{fig: continuum and line segments of Mo,both}
   \end{figure*}

\subsection{Population separation}
\label{sec: population sep}
The separation of the disk components is a combination of both chemical and kinematical properties. As described in more detail in \citetalias{lomaeva2019A&A...625A.141L}, we use [Ti/Fe] and [Fe/H] \citepalias[determined in][]{jonssona2017A&A...598A.100J} as a proxy for the chemical separation typically observed in $\alpha$-abundances. Additionally, we use Galactic space velocities as calculated with \texttt{galpy} \citep{galpybovy2015ApJS..216...29B} using radial velocities (see Table\,\ref{tab:basicdata_sn}), distances \citep{mcmillan:2018} and proper motions \citep{gaiacollaboration2016A&A...595A...1G,gaiacollab:2018} as input and calculate the total velocities, $V_{\mathrm{total}}^2 = U^2 + V^2 + W^2$. Since Gaia has a limit on brightness, some of our brightest stars are not observed with Gaia, and in total kinematic data were available for 268 of the disk stars. 
We then use the clustering method called \texttt{Gaussian Mixture Model} (GMM), which can be found in the \texttt{scikit-learn} module in Python \citep{pedregosa2011scikit}, to cluster the disk data into the two components. Since we use a combination of chemistry and kinematics, we refer to the components as thin- and thick disk, where the thick disk typically is more $\alpha$-rich and kinematically hotter than the thin disk.    

\section{Results}
\label{sec: results}
\begin{figure}[h!]
   \centering
   \includegraphics[width=\hsize]{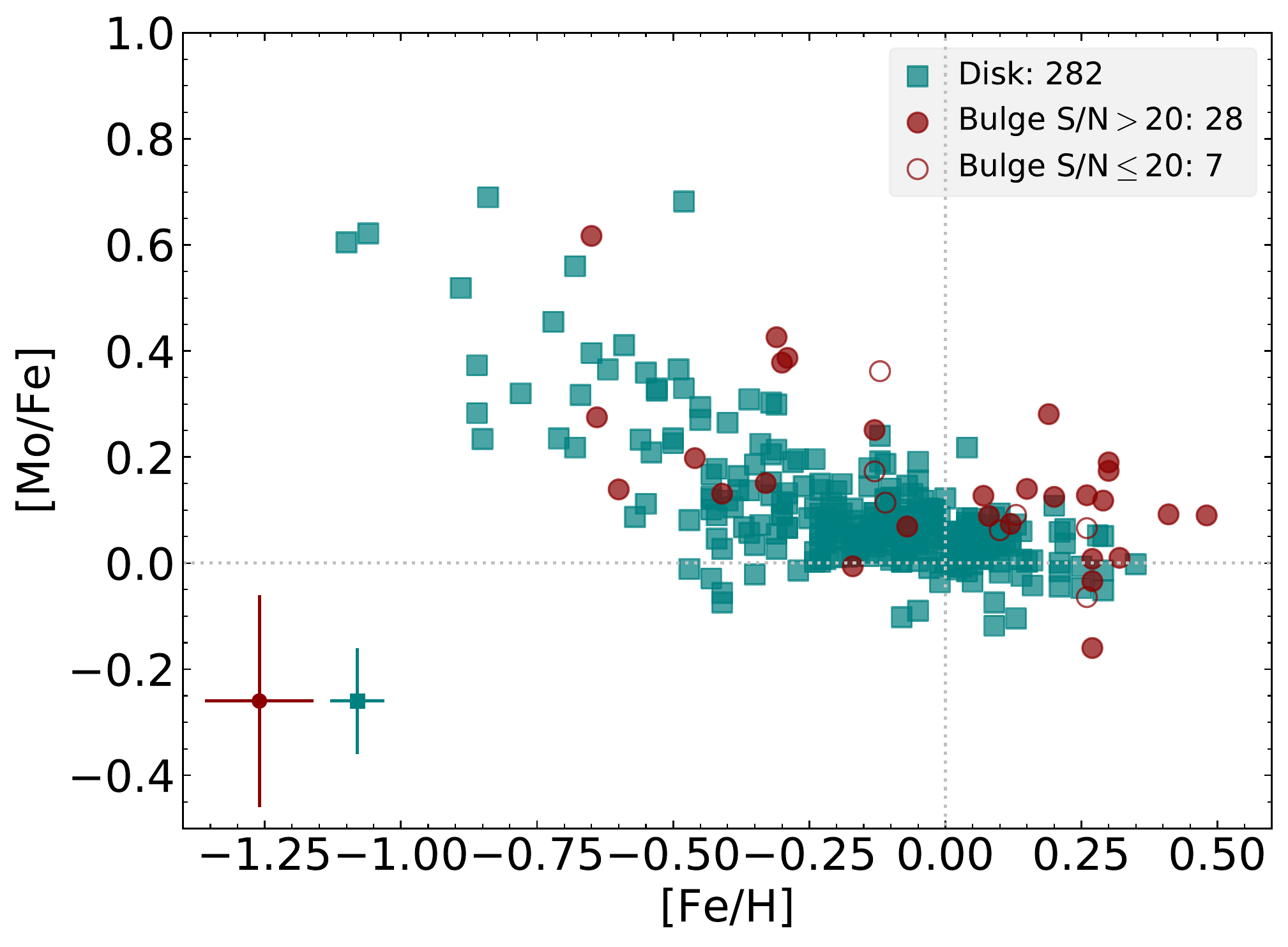}
      \caption{[Mo/Fe] vs [Fe/H] for the disk (teal) and the bulge (red) in this study. The typical uncertainties, as described in Section\,\ref{sec: uncertainties} are indicated in the lower right corner. The grey dashed lines that go through [0,0] indicate the solar value, which we have normalized to A(Fe) = 7.45 \citep{grevesse2007SSRv..130..105G} and A(Mo) = 1.88 \citep{grevesse2015A&A...573A..27G}.}
      \label{fig: Mo our work}
   \end{figure}

After the manual inspection of the synthetic spectra, we end up with 282 stars in the local disk and 35 in the bulge with reliably determined Mo abundances. The detailed abundances for the disk and bulge giants can be found in Table\,\ref{tab:abundances_sn} and \ref{tab:abundances_bulge}, respectively. The abundance ratio [Mo/Fe] plotted against the metallicity [Fe/H] for the stellar samples can be seen in Figure\,\ref{fig: Mo our work}. To be consistent with the previous work in this series, we have made a distinction between bulge spectra having a S/N of above or below 20. Nonetheless, as can be seen in Figure\,\ref{fig: Mo our work}, the bulge stars with S/N $\leq 20$ are within the scatter of the overall bulge trend. The typical uncertainties are noted in the lower left corner of the plot. The estimation of the uncertainties is described further in Section\,\ref{sec: uncertainties}.

\begin{figure}[h!]
   \centering
   \includegraphics[width=\hsize]{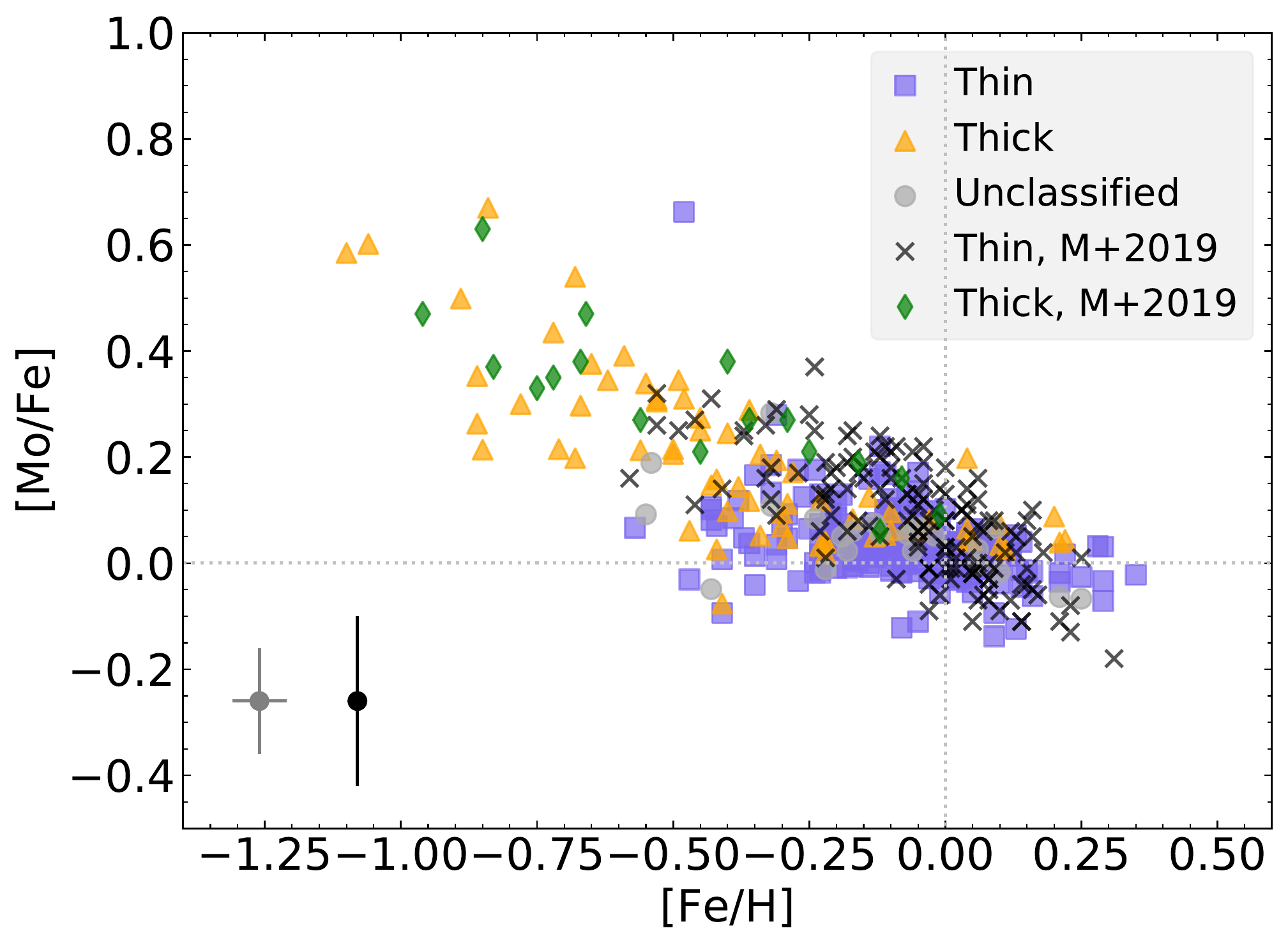}
      \caption{Comparing the [Mo/Fe] thin (blue squares) and thick (orange triangles) disk abundances determined in this work to the thin (black crosses) and thick (green diamonds) disk in \citet{mishenina2019MNRAS.489.1697M}. It should be noted that the 23 stars that we did not have kinematical data for have not been classified as either thin- or thick disk, and can be seen as grey circles. The typical uncertainties are indicated in the lower left corner (this work in grey, \citet{mishenina2019MNRAS.489.1697M} in black). The grey dashed lines that go through [0,0] indicate the solar value, which we have normalized to A(Fe) = 7.45 \citep{grevesse2007SSRv..130..105G} and A(Mo) = 1.88 \citep{grevesse2015A&A...573A..27G}. The thin disk star at [Fe/H] = -0.55 with high molybdenum abundance of [Mo/Fe] = 0.68, HIP65028, is discussed in the end of Section\,\ref{sec: discussion}.}
      \label{fig: thin thick ours Mishenina}
   \end{figure}

The separation of the disk components can be seen in Figure\,\ref{fig: thin thick ours Mishenina} where we also compare with disk sample from \citet{mishenina2019MNRAS.489.1697M}. With their work, they published the first extended sample of Mo-abundances for stars in the Milky Way disk, which, together with halo observations (as described below), help extend the knowledge of this element. With our additional sample, we now extend the study of Mo even more, and provide a comparison giant sample to the disk abundances. The \citet{mishenina2019MNRAS.489.1697M} sample consists of 183 disk stars, where they identify 163 as thin and 20 as thick disk dwarf stars \citep[determined using kinematics, ][]{mishenina2013A&A...552A.128M,mishenina2019MNRAS.489.1697M}, whereas we identify 191 and 68 thin- and thick disk giant stars. As such, we more than double and triple the thin and thick disk sample of Mo-abundances, respectively.

In the work of \citet{mishenina2019MNRAS.489.1697M} they too have high-resolution spectra of $\mathrm{R} > 42 000$ and $\mathrm{S/N} > 100$. In their abundance analysis of Mo I in their dwarf sample, they use the spectral lines at 5506 and 5533 Å. These two lines have quite strong blends, which can make abundance determination difficult \citep{heiter2021A&A...645A.106H}. Since these two lines are outside of the spectral range for our bulge stars, we have not included these in our analysis. Additionally, the 6030 Å line that is used in our study is not accessible in dwarf stars, where it is very weak (see Section\,\ref{sec: abu det Mo}).  



\begin{figure*}[h!]
   \centering
   \includegraphics[width=\hsize]{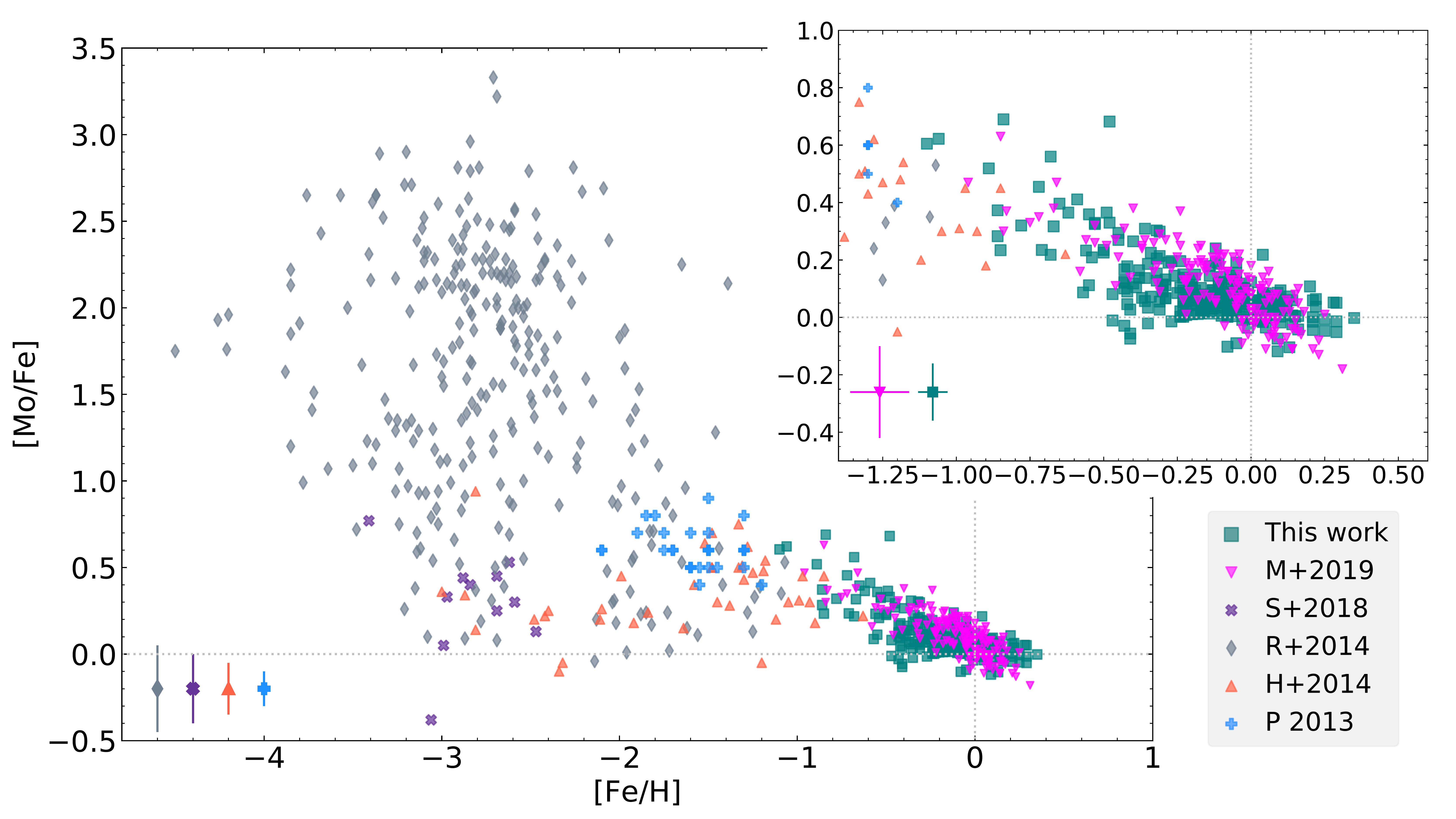}
      \caption{Comparing the [Mo/Fe] disk abundances determined in this work (teal squares, as in Figure\,\ref{fig: Mo our work}) to those in previous studies in the disk \citep[][magenta downward pointing triangles]{mishenina2019MNRAS.489.1697M} and the halo \citep[][blue pluses, gray diamonds, coral triangles, purple crosses, respectively]{peterson2013ApJ...768L..13P, roederer2014...313...AJ....147..136R, hansen2014A&A...568A..47H, spite2018A&A...611A..30S} This is also indicated in the lower right legend. The upper, rightmost, smaller plot shows a zoomed in portion of the higher metallicity region in the leftmost larger plot. The typical uncertainty reported in the studies are indicated in the lower left corner of the both plots. The grey dashed lines that go through [0,0] indicate the solar value, which we have normalized to A(Fe) = 7.45 \citep{grevesse2007SSRv..130..105G} and A(Mo) = 1.88 \citep{grevesse2015A&A...573A..27G}. 
      }
      \label{fig: Mo comp literature}
   \end{figure*}

In Figure\,\ref{fig: Mo comp literature} we plot some additional previous work in the more metal-poor regime of $\mathrm{[Fe/H]} < -1.2$, which consists mainly of halo stars. \citet{roederer2014...313...AJ....147..136R} uses the 3864 Å Mo I line to determine the Mo-abundances in both horizontal branch, main sequence, red- and subgiant stars. Altogether they determine Mo in 279 low-metallicity stars \citep[note that the two metal-poor stars in][are not included in Figure\,\ref{fig: Mo comp literature}]{roederer2014ApJ...791...32R}. It is worthwhile to note that even though the 3864 Å is not reachable in our sample which is limited to 5800-6800 Å, the bluer wavelength regime would be very crowded with lines in giant stars at the metallicities of our stellar sample, making continuum placement extremely difficult. 

In the work from \citet{hansen2014A&A...568A..47H} and \citet{peterson2013ApJ...768L..13P}, they primarily also use the 3864 Å Mo I line to determine Mo-abundances while \citet{peterson2013ApJ...768L..13P} focuses on turnoff stars, \citet{hansen2014A&A...568A..47H} has a combination of dwarfs and giant star, a total of 52. They have high quality data with spectral resolution of R $\sim$ 40 000 and S/N $>$ 100. Note that in Figure\,\ref{fig: Mo comp literature} we only plot the 40 stars that have abundances marked as high quality, and not affected by large uncertainties due to blends and continuum placement \citep[see Table 4 in][]{hansen2014A&A...568A..47H}. Finally, we also plot the Mo abundances of the 11 stars in \citet{spite2018A&A...611A..30S}, which also uses the bluer 3864 line. It should be noted that their star at [Fe/H] $- 3.06$ with [Mo/Fe] of -0.38 is a r-poor star, BD–18 5550, explaining the low abundance. 

There are also published Mo-abundances for barium stars (Ba-stars). These are stars enriched in s-process elements, as well as in carbon, but otherwise have nominal abundances. These stars have been enriched due to accretion of s-process elements from a companion AGB-star, resulting in these peculiar abundances \citep{allen2006A&A...454..917A, roriz2021MNRAS.507.1956R}. As such, we do not include Ba-stars in the comparison plot with previous Mo-abundances. Furthermore, Mo has been measured in the globular clusters M22 \citep{roederer2011ApJ...742...37R} and 47 Tucanae \citep{thygesen2014A&A...572A.108T}, in open clusters \citep[see e.g.][among others]{overbeek2016ApJ...824...75O, mishenina2020arXiv200604629M} and in the r-process enhanced bulge star 2MASS J18174532–3353235 in \citet{johnson2013ApJ...775L..27J}. Note that these are not included in Figure\,\ref{fig: Mo comp literature}, where we look at the overall disk (and halo) trend in molybdenum.

Lastly, in Figure\,\ref{fig: [Mo/Fe], [Ce/Fe], [Eu/Fe] thin, thick, bulge} and Figure\,\ref{fig: [Mo/Eu] and [Mo/Ce]} we compare the Mo-abundance with the abundances of Ce (s-process) and Eu (r-process) from \citetalias{Forsberg2019A&A...631A.113F}. We will discuss these Figures more in Section\,\ref{sec: discussion} below.
   
\subsection{Uncertainties}
\label{sec: uncertainties}
The random uncertainties that arise due to line- and continuum placements are hard to estimate, which also goes for the possible uncertainties in the atomic data and the model atmosphere assumptions used for the spectral line synthesis. As such, the uncertainties for the abundances determined in this work are deemed to be mostly affected by the possible uncertainties in the stellar parameters. 

The typical uncertainties for a local disk giant of the median S/N $\sim 100$ are estimated in \citetalias{jonssona2017A&A...598A.100J} to be on the order of $\mathrm{T}_{\mathrm{eff}} \pm 50 \ \mathrm{K}$, $\log g \pm 0.15 \ \mathrm{dex}$, $\mathrm{[Fe/H]} \pm 0.05 \ \mathrm{dex}$, and lastly $\pm \ 0.1 \ \mathrm{km/s}$ for $\xi_\mathrm{micro}$. Since the bulge stars have a general lower S/N, the uncertainties are estimated to be twice that of the disk stars \citepalias{jonsson2017b}. These values can be seen in the leftmost column of Table\,\ref{tab: uncertanties}, and are subsequently used to estimate the uncertainties for the Mo abundances.

To estimate the Mo-abundances, we add the uncertainties from \citetalias{jonssona2017A&A...598A.100J} to two typical giant stars, Arcturus (also known as $\alpha\mathrm{-Boo}$ or HIP69673) and Rasalas ($\mu\mathrm{-Leo}$ or HIP48455). We do this step-wise, and determine the abundance with that new set of stellar parameters, for both stars.

The total abundance uncertainties coming from the uncertainties in the stellar parameters are then calculated as
\begin{equation}
    \label{eq: un eq}
    \sigma \mathrm{[Mo/Fe]} = \sqrt{|\delta A_{T_{\text{eff}}}|^2 + |\delta A_{\log(\mathrm{g})}|^2 + |\delta A_{\text{[Fe/H]}}|^2 + |\delta A_{v_{\text{micro}}}|^2}
\end{equation}
where, for possible non-symmetrical abundance changes, the mean value is used in the squared sums. Taking the average of the $\delta$A(Mo)$_{\alpha\mathrm{-Boo}}$ and $\delta$A(Mo)$_{\mu\mathrm{-Leo}}$ as calculated from Eq.\,(\ref{eq: un eq}) we get a typical abundance uncertainty of 0.1 dex in the disk and 0.2 dex in the bulge. We list the total uncertainties in Table\,\ref{tab: uncertanties}. 

Stellar parameters in reality are coupled and change dependently on each-other, and this way of determining the uncertainties is a simplified approach. From Monte Carlo estimations of the uncertainties \citepalias[][]{lomaeva2019A&A...625A.141L,Forsberg2019A&A...631A.113F}, we find the uncertainties determined in this simplified way to yield very similar values. Additionally, considering the tight abundance trends we produce, the uncertainties can in general be considered to be upper limits.

\begin{table}[]
    \centering
    \caption{Table of estimated uncertainties for Mo using the typical giant stars $\alpha\mathrm{-Boo}$ and $\mu\mathrm{-Leo}$. The upper half of the table shows the typical uncertainties for a disk star, and the lower half for a typical bulge star. The total uncertainty, $\delta$[Mo/Fe]$_\mathrm{total}$, is calculated using Eq.\,(\ref{eq: un eq}). The values in the left column comes from \citetalias{jonssona2017A&A...598A.100J}.}
    \label{tab: uncertanties}
    \begin{tabular}{c| c c} \hline\hline
         Uncertainty & $\delta$[Mo/Fe]$_{\alpha\mathrm{-Boo})}$ & $\delta$[Mo/Fe]$_{\mu\mathrm{-Leo}}$\\  \hline \hline 
         $\delta$ \teff = -50 K & -0.05 & -0.05   \\
         $\delta$ \teff = +50 K & +0.05 & +0.05  \\
         $\delta$ \logg = -0.15 & -0.02 & -0.03   \\ 
         $\delta$ \logg = +0.15 & +0.03 & +0.03   \\ 
         $\delta$ [Fe/H] = -0.05 & -0.04 & +0.04   \\ 
         $\delta$ [Fe/H] = +0.05 & +0.04 & -0.04   \\ 
         $\delta$ \vmic = -0.10 & 0.00 & 0.00   \\ 
         $\delta$ \vmic = +0.10 & 0.00 & 0.00   \\ \hline
         
        $\delta$[Mo/Fe]$_\mathrm{total}$ Disk & 0.11 dex & 0.10 dex \\ \hline
         
         $\delta$ \teff = -100 K & -0.10 & +0.09   \\
         $\delta$ \teff = +100 K & +0.11 & +0.10  \\
         $\delta$ \logg = -0.30 & 0.00 & -0.05   \\ 
         $\delta$ \logg = +0.30 & +0.06 & +0.05   \\ 
         $\delta$ [Fe/H] = -0.10 & -0.08 & +0.08   \\ 
         $\delta$ [Fe/H] = +0.10 & +0.09 & -0.08   \\ 
         $\delta$ \vmic = -0.20 & +0.01 & +0.01   \\ 
         $\delta$ \vmic = +0.20 & 0.00 & -0.01   \\ \hline
         $\delta$[Mo/Fe]$_\mathrm{total}$ Bulge & 0.20 dex & 0.20 dex \\ \hline
    \end{tabular}
\end{table}

\section{Discussion}
\label{sec: discussion}
In the discussion, we will first go through and discuss the astrophysical sites for the s-, r- and p-processes. Then we compare with previous data of molybdenum, and end with a discussion of molybdenum as compared with other neutron-capture elements. Lastly, we also make a short comment on the star HIP65028.

\subsection{The neutron-capture and p-processes}
As mentioned in the introduction, Mo has a diverse origin, having stable isotopes originating from both the s-, r- and p-process, or a combination of these. As for the two neutron-capture processes, the sites where these take place are constrained via the required neutron flux.

The s-process can be divided into two sub-processes, the \textit{main} and the \textit{weak} s-process. The main s-process, takes place in the interior of low- and intermediate asymptotic giant branch (AGB) stars in a ${}^{13}\mathrm{C}$-pocket during the third dredge-up \citep[see][and references therein]{karakas2014PASA...31...30K,bisterzo2017ApJ...835...97B}. This process is the main s-process producer for elements with $\mathrm{A}\,\gtrsim 90$, such as cerium, which we will compare our Mo-abundances to. We refer the reader to \citetalias{Forsberg2019A&A...631A.113F} for a more in-depth description of the main s-process and its components.

The weak s-process requires higher temperatures and has the ${}^{22}\mathrm{Ne(}\alpha\mathrm{,n)}{}^{25}\mathrm{Mg}$-reaction as neutron source. As such, it takes place in the interior of massive stars with mass $\gtrsim 8$\,M$_{\odot}$. The weak s-process can produce trans-iron elements of $60 \lesssim  \mathrm{A} \lesssim 90$, making its affect on the production of molybdenum-isotopes likely very small and close to negligible \citep{jenniferjohnson2002,pignatari2010,prantzos020MNRAS.491.1832P}.

However, some studies \citep[e.g.][]{travaglio2004,bisterzo2014ApJ...787...10B,bisterzo2017ApJ...835...97B} find that an additional process, \textit{Light Element Primary Process} (LEPP), different from both the main and the weak s-process, would be necessary to explain the abundances of Sr, Y and Zr, as well as the s-only isotopes ${}^{96}\mathrm{Mo}$ and ${}^{130}\mathrm{Xe}$.

Nonetheless, the s-process has been proven difficult to model, due to its dependence on a wide range of physical parameters, and uncertainties of the yields \citep{cescutti2022Univ....8..173C}. The necessity of LEPP has also been questioned \citep[see e.g.][]{cristallo2011,cristallo2015,trippella,prantzos020MNRAS.491.1832P,kobayashi2020ApJ...900..179K} and the need for it has not been necessary when modifying parameters of Galactic Chemical Evolution models, such as the star formation rate, stellar yields, and varying the size of the ${}^{13}\mathrm{C}$-pocket. Additionally, the rotation of massive stars has also been shown to have an affect on the amount of s-process elements of $\mathrm{A} \lesssim 90$ being produced at low metallicities \citep{cescutti2013A&A...553A..51C,frischknecht2016MNRAS.456.1803F,limongi2018ApJS..237...13L}. Therefore, the s-process contribution to molybdenum is believed to mainly come from the main s-process in AGB-stars.

Given the AGB-origin of main s-process elements, we expect these to have a trailing end at lower metallicities in abundance plots, caused by the natural delay-time of AGB-stars. As AGB stars starts enriching the ISM with s-process elements, the abundance of these elements will increase in newly formed stars, resulting in an increase of [s/Fe], before SNe type Ia starts enriching with iron, bringing the abundance trend down again \citep[see e.g. observations in][]{mishenina2013A&A...552A.128M,battistini2016,delgadomena,Forsberg2019A&A...631A.113F}.

The r-process produces elements such as Eu, which we also will compare Mo to. Given the high neutron-flux required for the r-process, this is a production channel that works on very short time scales. The proposed production sites for the r-process are various SNe such as core-collapse \citep[core collapse, magneto-rotational, electron capture,][]{woosleyrprocess1994ApJ...433..229W,nishimura2006ApJ...642..410N,kobayashi2020ApJ...900..179K,wanajo2011ApJ...726L..15W} and neutron star mergers \citep[NMS,][]{Freiburghaus1999ApJ...525L.121F, matteucci2014MNRAS.438.2177M}. In observations of the electromagnetic signature from the NSM GW170817 \citep{abbott2017PhRvL.119p1101A}, r-process ejecta could be detected \citep{abbott2017ApJ...848L..12A,kasen2017Natur.551...80K}. However,  \citet{kobayashi2020ApJ...900..179K} uses Galactochemical evolution models to show that NSM are more or less negligible in the production of r-process elements, and point to MRSNe as the major contributor. \citet{cote2019ApJ...875..106C} and \citet{skuladottir2020A&A...634L...2S} points out the necessity of having a combination of two sources with different delay-times in order to reproduce observed abundance trends, which is very similar to that of $\alpha$-elements, in the Milky Way and some of the dwarf galaxies. To conclude, there are still uncertainties to the amount of contribution from different sources. To determine the contribution from the suggested production sites is an active area of research, and having high-quality observational data to match models against are key in this continued effort.

The site of the p-process is even less understood. It was suggested in \citet{cameron1957PASP...69..201C, burbidge1957} to take place in hydrogen-rich layers of type II SNe. The name "p-process" refers to proton-capture, however this is not necessarily always the case and there are several mechanisms and sites suggested as the cosmic origin for the p-isotopes, which we will outline here \citep[see also the review of][]{rauscher2013RPPh...76f6201R}.

\textbf{The $\gamma$-process} is the photo-disintegration by highly energetic gamma-photons of neutrons in heavy, neutron-rich isotope which have already been created by the means of neutron-capture processes \citep{woosley1978ApJS...36..285W,arnould2003PhR...384....1A, hayakawa2004PhRvL..93p1102H, pignatari2016IJMPE..2530003P}. The $\gamma$-process and has been proposed to take place in the explosive O/Ne-shell burning stages of core-collapse SNe  \citep{hoffman1994AAS...185.3309H,rayet1995A&A...298..517R,  pignatari2016IJMPE..2530003P, nishimura2018MNRAS.474.3133N, travaglio2018ApJ...854...18T}, in pre-SNe stages \citep{ritter2018MNRAS.474L...1R} or in SNe type Ia \citep{howard1991ApJ...373L...5H, travaglio2011ApJ...739...93T,travaglio2015ApJ...799...54T}. As such, we expect a signature from these processes to be similar to the signatures for $\alpha-$ and iron-peak elements, also originating from core-collapse SNe and SNe type Ia, respectively.



\textbf{The $\nu$-process}, that takes place during core-collapse SNe, could also be responsible for the production of p-isotopes. This process has been in favour in making some of the lightest p-isotopes, including ${}^{92,94}\mathrm{Mo}$ \citep{woosley1990ApJ...356..272W, froclich2006PhRvL..96n2502F}. 

\textbf{The rapid proton-capture processes} can be divided into three different processes, the \textit{rp-, pn-} and \textit{$\nu$p-process}. The rapid proton capture takes place in proton rich environments through (p,$\gamma$)-reactions \citep{schatz1998PhR...294..167S}. 
The rp-process is usually halted at isotopes having a combination of long half-lives and small proton-capture cross sections, causing "waiting points" \citep{schatz2001PhRvL..86.3471S}. The \textit{rp-process} has been suggested to be linked to the explosive hydrogen- and helium-burning on the surface of mass-accreting neutron stars \citep{schatz1998PhR...294..167S,koike2004ApJ...603..242K}.

The waiting points which halts the rp-process can be crossed either by the \textit{pn}- or the \textit{$\nu$p-process}. If there is a high density of neutrons, (n,p)-reactions can cross the waiting points, which is the pn-process. Otherwise, neutrons can be created through anti-neutrino interactions with protons, in $\overline{\nu}_\mathrm{e}\mathrm{(p,e}^+\mathrm{)n}$-reactions, which is the \textit{$\nu$p-process} \citep{froclich2006PhRvL..96n2502F}. 

The pn-process has been suggested to take place in a subclass of type Ia supernovae which is caused by the disruption of a sub-Chandrasekhar CO-white dwarfs due to a thermonuclear runaway in He-rich accretion layers \citep{goriely2002A&A...383L..27G}. 
The site for the $\nu$p-process has been suggested to be either core-collapse SNe explosions or at accretion disks around compact objects \citep{froclich2006PhRvL..96n2502F,frohlich2017nuco.confa0505F,eichler2018JPhG...45a4001E}. 

In conclusion, there is a plethora of different mechanisms and sites that could produce the p-isotopes. By comparing to neutron-capture elements that dominate in the s- and r-process, the p-process might be disentangled from these.

\subsection{Comparison with previous work on molybdenum}
As noted earlier, the disk [Mo/Fe]-trend in \citet{mishenina2019MNRAS.489.1697M} decrease at higher metallicities, whereas our disk trend decrese to a lesser degree. Furthermore, considering the separation of the thin and thick disk in Figure\,\ref{fig: thin thick ours Mishenina}, we see that the solar-[Mo/Fe] found at around [Fe/H] $\sim -0.4$ in our data, corresponds to the thin disk, which is not observed in \citet{mishenina2019MNRAS.489.1697M}. 
As such, our [Mo/Fe]-trend with metallicity for the thin disk have more of a trailing end at lower metallicities, which is typical for an s-process element that has a dominating enrichment from AGB-stars at $\mathrm{[Fe/H]} \gtrsim -0.5$, as discussed above. 

As can be seen in Figure\,\ref{fig: Mo comp literature}, the cosmic scatter at $\mathrm{[Fe/H]} < -1.5$ is extremely large, mostly based on the data from \citet{roederer2014...313...AJ....147..136R}. Similar large spread at low metallicities is also seen for other neutron-capture elements, such as Eu \citep{francois2007A&A...476..935F,frebel2010AN....331..474F,cescutti2015A&A...577A.139C} and ytterbium \citep[Yb, Z = 70, also as published in][]{roederer2014...313...AJ....147..136R} whereas at higher metallicities tighter trends are observed \citep[see e.g.][for Yb]{montelius2022arXiv220200691M}. The stochastic Galactochemical evolution models from \citet{cescutti2015A&A...577A.139C} can reproduce these observations and show that neutron-capture elements indeed have a large spread at lower metallicities. This is explained by the fact that at low metallicities, all neutron-capture elements in fact have r-process origin, since the onset of s-process production of AGB has not taken place yet. The spread could be a signature of local r-process production, such as NSM, as discussed in \citet{cescutti2015A&A...577A.139C}. This, in combination with the Galaxy not being well-mixed, and indeed very low in Fe, could give rise to locally high [r-process/Fe], which is observed for [Mo/Fe].

It should be noted that the scatter at the lower metallicities, spanning between $\mathrm{[Mo/Fe]} \sim 0 - 3.3$ dex mostly originates from the data of \citet{roederer2014...313...AJ....147..136R}, whereas the abundances of \citet{peterson2013ApJ...768L..13P,hansen2014A&A...568A..47H,spite2018A&A...611A..30S} are found within $\mathrm{[Mo/Fe]} \sim 0 - 1$ dex. More abundances for these low-metallicity stars would be needed to further constrain the physical origin of this scatter for the r-process.

\begin{figure*}[h!]
   \centering
   \includegraphics[width=\hsize]{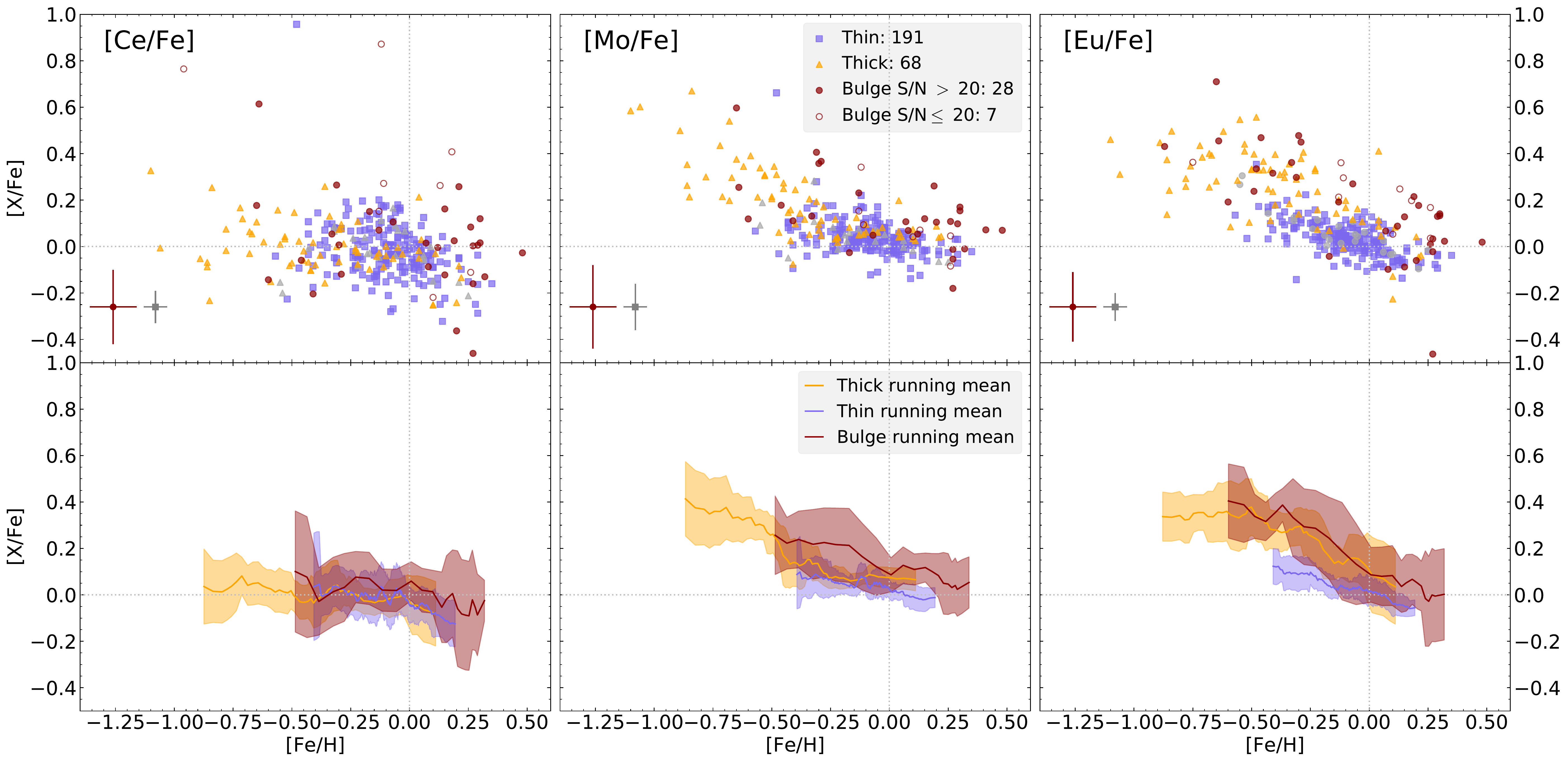}
      \caption{[Mo/Fe] over [Fe/H] for the thick disk (yellow), thin disk (blue) and bulge (red) stars seen in the middle panels. The number of stars in each population are marked in the legend of the scatter plot (upper). The running mean (lower) is calculated using a box size of 10 (thick), 20 (thin) and 7 (bulge) stars and plotted with a 1$\sigma$ deviation. Note that for the bulge we only use stars with a spectra of $> 20$ S/N to produce the running mean. We compare it to the s-process element Ce (left) and r-process element Eu (right) from \citet{Forsberg2019A&A...631A.113F}. The typical uncertainties are indicated in the lower right corner of the plots (red for bulge, grey for disk). The grey dashed lines that go through [0,0] indicate the solar value, which we have normalized to A(Fe) = 7.45, A(Ce) = 1.70 \citep{grevesse2007SSRv..130..105G}, A(Mo) = 1.88, A(Eu) = 0.52 \citep{grevesse2015A&A...573A..27G}. The [Eu/Fe] values in \citet{Forsberg2019A&A...631A.113F} are reported to likely be systematically too high, possibly originating from the systematic uncertainties in $\log(g)$ as reported in \citetalias{jonssona2017A&A...598A.100J}. Due to this, we lower the [Eu/Fe] abundances by 0.10\,dex in this figure, such that the thin disk in [Eu/Fe] goes through the solar value.}
    \label{fig: [Mo/Fe], [Ce/Fe], [Eu/Fe] thin, thick, bulge}
   \end{figure*}

\subsection{Comparison with other neutron-capture elements}
We investigate the origin of Mo by comparing it to Ce which has a roughly $85\ \%$ contribution from the s-process \citep{bisterzo2014ApJ...787...10B,prantzos020MNRAS.491.1832P} and to Eu which has a roughly $95\ \%$ contribution from the r-process \citep{bisterzo2014ApJ...787...10B,prantzos020MNRAS.491.1832P}, making these elements key representatives of these processes in the Galactic stellar populations. The abundance data for Ce and Eu in the disk and bulge come from \citetalias{Forsberg2019A&A...631A.113F}, and can be seen in Figure\,\ref{fig: [Mo/Fe], [Ce/Fe], [Eu/Fe] thin, thick, bulge}. As can be seen, the down turning trailing end at lower metallicities is more prominent in [Ce/Fe] than in [Mo/Fe], which is expected from the theoretical origins of these elements. Moreover, both the thick disk and the bulge show decreasing [Mo/Fe]-trends with increasing metallicity, which is typical for a r-process element, due to the fast enrichment \citep[see the models in][]{matteucci2014MNRAS.438.2177M,grisoni2020MNRAS.492.2828G}. 

In [Eu/Fe] we observe a 'knee', or plateau in the thick disk at $\mathrm{[Fe/H]} \leq -0.5$, which is not observed for [Mo/Fe]. It does not seem to plateau at all. However, more abundances at these metallicites are needed to further constrain this. With the combined data from \citet{peterson2013ApJ...768L..13P,hansen2014A&A...568A..47H,roederer2014...313...AJ....147..136R} it does seem like [Mo/Fe] has a knee around $\mathrm{[Fe/H]} \sim -0.75$, before the, as previously discussed, increased scatter at low metallicities. 

As previously stated in \citetalias{Forsberg2019A&A...631A.113F}, the general similarity between the thick disk and the bulge is high in [Ce, Eu/Fe], as seen in the running means of the abundances in the second row of Figure\,\ref{fig: [Mo/Fe], [Ce/Fe], [Eu/Fe] thin, thick, bulge}, pointing at a similar star formation history of the populations. The [Ce,Eu/Fe] abundances in the running means are at some metallicities $\sim 0.1$ dex higher in the bulge than in the thick disk, however as the scatter is large in the bulge, we refrain from drawing any hard conclusions.

It seems like the bulge possibly is enriched in [Mo/Fe], compared to the thick disk, especially at $-0.5 < \mathrm{[Fe/H]} < 0$. However, we do note that the scatter is large for the bulge, and the following discussion is speculative based on the possible enrichment. Since we do not see any difference between the stellar populations in either [Ce/Fe] or [Eu/Fe], the high [Mo/Fe] in the bulge could be linked to the p-process. In our previous studies, we found a possible enrichment in [V, Co, La/Fe] for the bulge, but these differences are not consistent with a single type of nucleosynthetic origin, meaning that we — all elements considered — cannot with confidence say that the bulge is distinguishable from the thick disk in the amount and rate of SNe type I and II. In turn, this means that the p-process sites linked to SNe type II and WD-sources likely cannot produce the possible difference in the thick disk and bulge [Mo/Fe]. Out of the suggested p-process sites, the one remaining then is the $rp$-process, linked to burning on mass-accreting neutron stars \citep[as discussed above,][]{schatz1998PhR...294..167S,koike2004ApJ...603..242K}. However, there is no obvious reason why such events would be more common in the bulge, since either a high star formation rate or top-heavy initial mass function would make the bulge stand out in $\alpha$-elements.

   \begin{figure}[h!]
   \centering
   \includegraphics[width=\hsize]{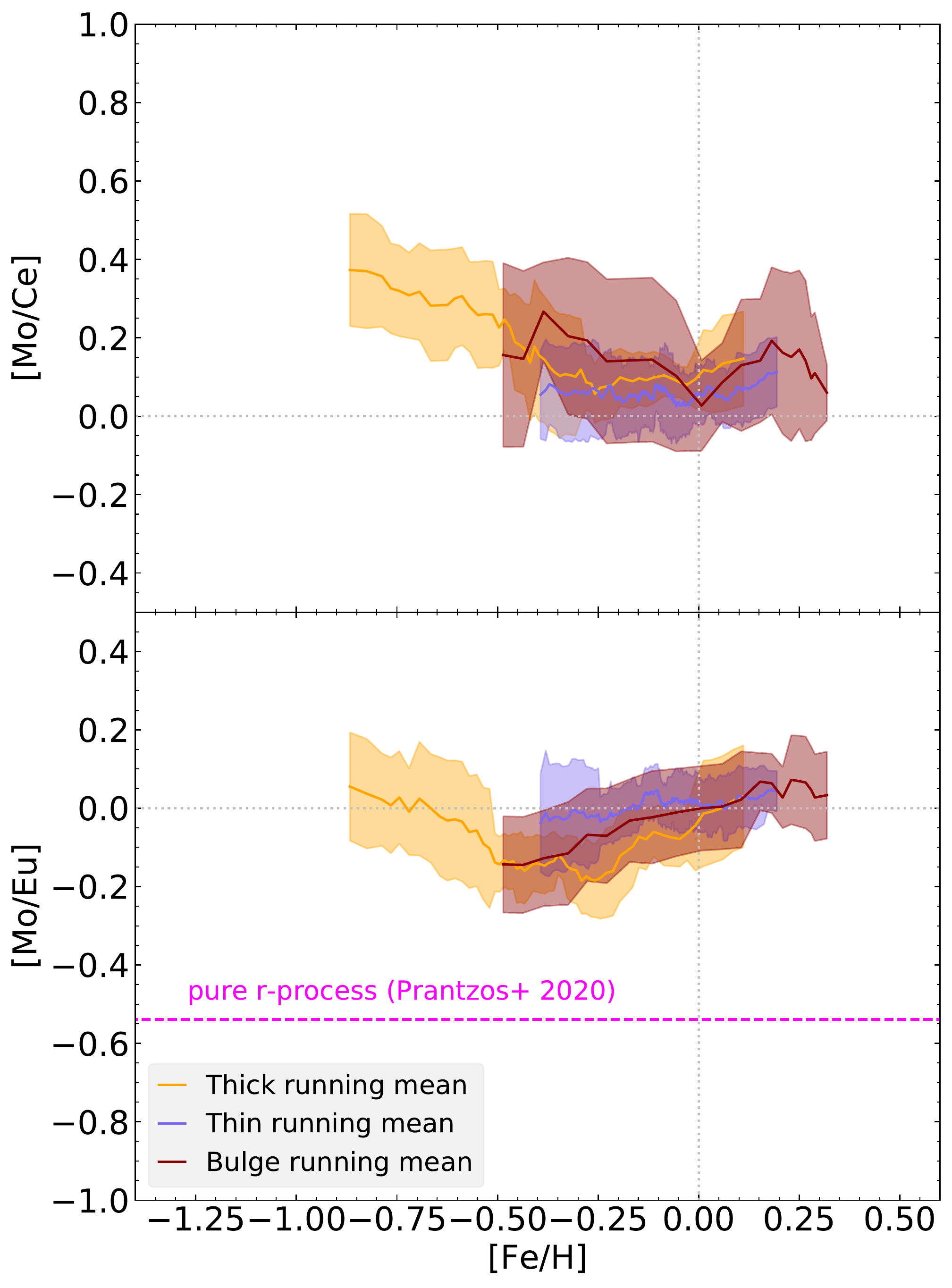}
      \caption{Comparison of Mo with the s-process element Ce (top) and the r-process element Eu (bottom), for the running means of the populations, similarly as Figure\,\ref{fig: [Mo/Fe], [Ce/Fe], [Eu/Fe] thin, thick, bulge}. We have also plotted the pure r-process line using the values in \citet{prantzos020MNRAS.491.1832P} (magenta).}
         \label{fig: [Mo/Eu] and [Mo/Ce]}
   \end{figure}

In order to further investigate the s- and r-process contribution to Mo, we also consider the abundance ratios of [Mo/Ce] and [Mo/Eu] over metallicity in Figure\,\ref{fig: [Mo/Eu] and [Mo/Ce]}. Here we only plot the running means, to help with the interpretation of the trends. We also plot the so-called "pure r-process line" in the [Mo/Eu]-plot, which is calculated using the solar r-process contributions of the elements. The pure r-process is then value of the r-process contribution in both elements, [r-process(Mo)/r-process(Eu)]. The closer the ratio is to the pure-r-process line, the more of the elements have a joint origin from the r-process (or rather are enriched by a source having similar time- and enrichment rate).

Since all neutron-capture elements are produced by the r-process at lower metallicities, we can see that Mo has a higher production from the r-process, compared to Ce, at low metallicities $\leq -0.5$. At $\mathrm{[Fe/H]} = -0.5$, AGB-stars (originating from low- to intermediate-mass) start to enrich the Galactic ISM with s-process elements, producing a higher fraction of Mo and Ce than the r-process. As a result, the [Mo/Ce]-trend flattens out.  
Considering [Mo/Eu] in Figure\,\ref{fig: [Mo/Eu] and [Mo/Ce]}, we see an increase with increasing metallicities, again at $\mathrm{[Fe/H]} = -0.5$, due to the AGB-enrichment, pointing at the s-process contribution to Mo. Below those metallicities, we observe a relative increase in Mo over Eu, which is linked to the lack of 'knee'/plateau in Mo at these metallicities. 


\subsection{Comment on HIP65028}
The star HIP65028 stands out as an Mo-enriched star in the thin disk, with [Mo/Fe] = 0.68. Considering the values for other neutron-capture elements, [La/Fe] = 1.04, [Ce/Fe] = 0.96, [Eu/Fe] = 0.37 \citepalias{Forsberg2019A&A...631A.113F}, we can conclude that this is a star enriched in s-process elements rather than r-process elements. This makes it a candidate for being a Barium star, which could be investigated further by determining its C-abundance (which in Ba-stars usually is high) or if it has a strong variance in radial velocity measurements, which would indicate a binary companion. Usually these companions are WDs, stripped AGB-stars from where the carbon and s-process elements originate, and are too faint to be directly observed. Nonetheless, this also shows the high contribution from the s-process in creating Mo \citep[$\sim$ 40 to 50 \% in][, respectively]{bisterzo2014ApJ...787...10B,prantzos020MNRAS.491.1832P} even at the low metallicities of [Fe/H] $\sim -0.5$ dex.

\section{Conclusions}
\label{sec: conclusions}
In this work we continue the series of determining high-quality abundances of bulge and disk stars, using high-resolution, optical spectra, with the aim of making a differential comparison between the stellar populations. We use the same type of stars (K-giants), the same spectral lines and atomic data, as well as the same method for analysing and determining the Mo-abundances. By using the same set of spectral lines for all stars, we minimize the possible systematic uncertainties in the comparison of the bulge and the disk.

We determine Mo-abundances for 35 bulge stars, which to the best of our knowledge, is the largest sample of Mo-abundances in bulge stars. We also determine Mo for 282 disk stars. With the previous sample of 183 disk stars from \citet{mishenina2019MNRAS.489.1697M}, our sample increases significantly the Mo-abundances determined at  $\mathrm{[Fe/H]} \gtrsim -1.2$. This allows us to both compare the bulge and disk stellar populations, and to investigate the cosmic origin of Mo.

Mo has an interesting cosmic origin, being composed of a combination of seven stable isotopes, all with s-, r- or/and p-process origin. We can not determine the abundance of specific isotopes, but the overall abundance of Mo in our stellar populations. In comparison with other neutron-capture elements, e.g. Ce and Eu, we can not exclude that Mo seems to stand out in the bulge, as compared to the thick disk. This could possibly be explained by the roughly $25 \ \%$ p-process origin of Mo (at solar metallicities). As such, the explanation to this possible difference between the bulge and thick disk could lie in understanding how and where these p-isotopes, namely ${}^{92,94}\mathrm{Mo}$, form. The production site of p-isotopes is yet to be fully constrained, and even though the scatter in our bulge sample is large, we speculate that an origin through mass-accreting neutron-stars could possibly explain the seemingly high bulge abundances.

\begin{acknowledgements}
We thank Ross Church for insightful discussions about neutron-capture elements. We also thank the anonymous referee for insightful comments that helped to improve this paper. R.F. and N.R. acknowledge support from the Royal Physiographic Society in Lund through the Stiftelse Walter Gyllenbergs fond and Märta och Erik Holmbergs donation. R.F's and A.J's research is supported by the Göran Gustafsson Foundation for Research in Natural Sciences and Medicine. A.J. acknowledges funding from the European Research Foundation (ERC Consolidator Grant 724687-PLANETESYS), the Knut and Alice Wallenberg Foundation (Wallenberg Academy Fellow Grant 2017.0287) and the Swedish Research Council (Project Grant 2018-04867). 
This work has made use of data from the European Space Agency (ESA) mission {\it Gaia} (\url{https://www.cosmos.esa.int/gaia}), processed by the {\it Gaia} Data Processing and Analysis Consortium (DPAC,
\url{https://www.cosmos.esa.int/web/gaia/dpac/consortium}). Funding for the DPAC as been provided by national institutions, in particular the institutions participating in the {\it Gaia} Multilateral Agreement. 
\textit{Software:} \texttt{matplotlib} \citep{matplotlib2007CSE.....9...90H}, \texttt{galpy} \citep[\url{http://github.com/jobovy/galpy},][]{galpybovy2015ApJS..216...29B}, \texttt{astropy} \citep{astropy2013A&A...558A..33A} .
\end{acknowledgements}

\bibliographystyle{aa} 
\bibliography{references.bib} 

\begin{appendix}
\section{Supplementary data}

\begin{table*}[h!]
\caption{Basic data for the observed solar neighbourhood giants. Coordinates and magnitudes are taken from the SIMBAD database, while the radial velocities are measured from the spectra. The S/N per data point is measured by the IDL-routine \texttt{der\textunderscore snr.pro}, see \href{http://www.stecf.org/software/ASTROsoft/DER\textunderscore SNR}{http://www.stecf.org/software/ASTROsoft/DER\textunderscore SNR}.}
\begin{tabular}{l l l l r r r l}
\hline
\hline
HIP/KIC/TYC & Alternative name & RA (J2000) & Dec (J2000) & $V$ & $v_{\mathrm{rad}}$ & S/N &  Source\\
        &                  & (h:m:s)    & (d:am:as)   &     & km/s\\
\hline
HIP1692 &           HD1690 & 00:21:13.32713 & $-$08:16:52.1625 &  9.18 &   18.37 &  114 & FIES-archive \\
HIP9884 &           alfAri & 02:07:10.40570 & +23:27:44.7032 &  2.01 &  $-$14.29 &   90 & PolarBase \\
HIP10085 &          HD13189 & 02:09:40.17260 & +32:18:59.1649 &  7.56 &   26.21 &  156 & FIES-archive \\
HIP12247 &            81Cet & 02:37:41.80105 & $-$03:23:46.2201 &  5.66 &    9.34 &  176 & FIES-archive \\
HIP28417 &          HD40460 & 06:00:06.03883 & +27:16:19.8614 &  6.62 &  100.64 &  121 & PolarBase \\
HIP33827 &           HR2581 & 07:01:21.41827 & +70:48:29.8674 &  5.69 &  $-$17.99 &   79 & PolarBase \\
HIP35759 &          HD57470 & 07:22:33.85798 & +29:49:27.6626 &  7.67 &  $-$30.19 &   85 & PolarBase \\
HIP37447 &           alfMon & 07:41:14.83257 & $-$09:33:04.0711 &  3.93 &   11.83 &   71 & Thygesen et al. (2012)\\
HIP37826 &           betGem & 07:45:18.94987 & +28:01:34.3160 &  1.14 &    3.83 &   90 & PolarBase \\
HIP43813 &           zetHya & 08:55:23.62614 & +05:56:44.0354 &  3.10 &   23.37 &  147 & PolarBase \\
\hline
\label{tab:basicdata_sn}
\end{tabular}
\tablefoot{This is only an excerpt of the table to show its form and content. The complete table is available in electronic form at the CDS.}
\end{table*}

\begin{table}[h!]
\caption{Stellar parameters and determined abundances for observed solar neighbourhood giants. [Fe/H] is listed in the scale of \citet{grevesse2007SSRv..130..105G}.}
\begin{tabular}{l c c c c c c}
\hline
\hline
HIP/KIC/TYC & $T_{\mathrm{eff}}$ & $\log g$ & [Fe/H] & $v_{\mathrm{micro}}$ & A(Mo)\\ 
\hline
HIP1692  & 4216 & 1.79 & $-$0.26 & 1.55 & 1.64 \\
HIP9884  & 4464 & 2.27 & $-$0.21 & 1.34 & 1.74 \\
HIP10085 & 4062 & 1.44 & $-$0.32 & 1.63 & 1.65 \\
HIP12247 & 4790 & 2.71 & $-$0.04 & 1.40 & 1.94 \\
HIP28417 & 4746 & 2.56 & $-$0.25 & 1.40 & 1.64 \\
HIP33827 & 4235 & 1.99 &    0.01 & 1.50 & 1.95 \\
HIP35759 & 4606 & 2.47 & $-$0.15 & 1.42 & 1.80 \\
HIP37447 & 4758 & 2.73 & $-$0.04 & 1.35 & 1.91 \\
HIP37826 & 4835 & 2.93 &    0.07 & 1.24 & 2.04 \\
HIP43813 & 4873 & 2.62 & $-$0.07 & 1.51 & 2.00 \\
\hline
\label{tab:abundances_sn}
\end{tabular}
\tablefoot{This is only an excerpt of the table to show its form and content. The complete table is available in electronic form at the CDS.}
\end{table}

\begin{table}[h!]
\caption{Basic data for the observed bulge giants. The S/N per data point is measured by the IDL-routine \texttt{der\textunderscore snr.pro}, see \href{http://www.stecf.org/software/ASTROsoft/DER_SNR}{http://www.stecf.org/software/ASTROsoft/DER\textunderscore SNR}.}
\begin{tabular}{l c c c c}
\hline
\hline
Star$^a$ & RA (J2000) & Dec (J2000) & $V$ & S/N\\
         & (h:m:s)    & (d:am:as)   &     & \\\hline
SW-09 & 17:59:04.533 & $-$29:10:36.53 & 16.153 & 16\\
SW-15 & 17:59:04.753 & $-$29:12:14.77 & 16.326 & 15\\
SW-17 & 17:59:08.138 & $-$29:11:20.10 & 16.388 & 11\\
SW-18 & 17:59:06.455 & $-$29:10:30.53 & 16.410 & 14\\
SW-27 & 17:59:04.457 & $-$29:10:20.67 & 16.484 & 13\\
SW-28 & 17:59:07.005 & $-$29:13:11.35 & 16.485 & 16\\
SW-33 & 17:59:03.331 & $-$29:10:25.60 & 16.549 & 14\\
SW-34 & 17:58:54.418 & $-$29:11:19.82 & 16.559 & 12\\
SW-43 & 17:59:04.059 & $-$29:13:30.26 & 16.606 & 16\\
SW-71 & 17:58:58.257 & $-$29:12:56.97 & 16.892 & 14\\
\hline
\end{tabular}
\label{tab:basicdata_bulge}
\tablefoot{This is only an excerpt of the table to show its form and content. The complete table is available in electronic form at the CDS.\\
\tablefootmark{a}{Using the same naming convention as \citet{lecureur2007} for the B3-BW-B6-BL-stars.}
}
\end{table}

\begin{table}[h!]
\caption{Stellar parameters and determined abundances for observed bulge giants. [Fe/H] is listed in the scale of \citet{grevesse2007SSRv..130..105G}.}
\begin{tabular}{l c c c c c c}
\hline
\hline
Star & $T_{\mathrm{eff}}$ & $\log g$ & [Fe/H] & $v_{\mathrm{micro}}$ & A(Mo) \\ 
\hline
SW-09 & 4095 & 1.79 & $-$0.15 & 1.32 & 1.90\\
SW-15 & 4741 & 1.96 & $-$0.98 & 1.62 & ...\\
SW-17 & 4245 & 2.09 &    0.24 & 1.44 & 2.06\\
SW-18 & 4212 & 1.67 & $-$0.13 & 1.49 & 1.86\\
SW-27 & 4423 & 2.34 &    0.11 & 1.60 & 2.08\\
SW-28 & 4254 & 2.36 & $-$0.14 & 1.44 & 2.10\\
SW-33 & 4580 & 2.72 &    0.16 & 1.39 & ...\\
SW-34 & 4468 & 1.75 & $-$0.45 & 1.63 & ...\\
SW-43 & 4892 & 2.34 & $-$0.77 & 1.84 & ...\\
SW-71 & 4344 & 2.66 &    0.39 & 1.31 & 2.36\\
\hline
\label{tab:abundances_bulge}
\end{tabular}
\tablefoot{This is only an excerpt of the table to show its form and content. The complete table is available in electronic form at the CDS.}
\end{table}

\end{appendix}

\end{document}